\documentclass[conference]{IEEEtran}
\IEEEoverridecommandlockouts

\usepackage[prologue,table]{xcolor}

\usepackage[utf8]{inputenc}
\usepackage{graphicx}
\usepackage{xspace}
\usepackage{tabularx}
\usepackage[]{hyperref}
\usepackage{listings}
\usepackage{booktabs}
\usepackage{rotating}
\usepackage[compact]{titlesec}
\usepackage[font=small,skip=0pt]{caption}
\usepackage[inline]{enumitem}
\usepackage[autolanguage]{numprint}
\usepackage[framemethod=tikz]{mdframed}
\usepackage[linesnumbered,noend]{algorithm2e}
\usepackage{slashbox}
\usepackage{amsmath}
\newtheorem{metric}{Metric}

\def\BibTeX{{\rm B\kern-.05em{\sc i\kern-.025em b}\kern-.08em
    T\kern-.1667em\lower.7ex\hbox{E}\kern-.125emX}}

\mdfdefinestyle{mpdframe}{
    frametitlebackgroundcolor   =black!15,
    frametitlerule              =true,
    roundcorner                 =3pt,
    middlelinewidth             =0.8pt,
    innermargin                 =0.1cm,
    outermargin                 =0.1cm,
    innerleftmargin             =0.1cm,
    innerrightmargin            =0.1cm,
    innertopmargin              =0.1cm,
    innerbottommargin           =0.1cm
}

\newcommand{\conform}{\texttt{Conforming}\xspace}  
\newcommand{\silent}{\texttt{Silent}\xspace}  
\newcommand{\error}{\texttt{Error}\xspace}

\newcommand{\json}{JSON\xspace}

\colorlet{punct}{red!60!black}
\definecolor{background}{HTML}{EEEEEE}
\definecolor{delim}{RGB}{20,105,176}
\colorlet{numb}{magenta!60!black}

\lstdefinelanguage{json}{
    basicstyle=\scriptsize\normalfont\ttfamily,
    numbers=left,
    numberstyle=\scriptsize,
    stepnumber=1,
    numbersep=8pt,
    showstringspaces=false,
    breaklines=true,
    literate=
     *{0}{{{\color{numb}0}}}{1}
      {1}{{{\color{numb}1}}}{1}
      {2}{{{\color{numb}2}}}{1}
      {3}{{{\color{numb}3}}}{1}
      {4}{{{\color{numb}4}}}{1}
      {5}{{{\color{numb}5}}}{1}
      {6}{{{\color{numb}6}}}{1}
      {7}{{{\color{numb}7}}}{1}
      {8}{{{\color{numb}8}}}{1}
      {9}{{{\color{numb}9}}}{1}
      {:}{{{\color{punct}{:}}}}{1}
      {,}{{{\color{punct}{,}}}}{1}
      {\{}{{{\color{delim}{\{}}}}{1}
      {\}}{{{\color{delim}{\}}}}}{1}
      {[}{{{\color{delim}{[}}}}{1}
      {]}{{{\color{delim}{]}}}}{1},
}

\usepackage{listings}

\usepackage{fp}
\usepackage{xspace}
\usepackage{numprint}
\newcommand*\np[2][z]{
\ifx z#1%
$\numprint{#2}$%
\else%
$\numprint[#1]{#2}$%
\fi\xspace
}

\newcommand{\ShowAbsoluteNumber}[1]{%
\ifnum #1<10%
{\hspace*{0pt}#1}%
\else%
\ifnum #1<100%
{\hspace*{0pt}#1}%
\else%
\ifnum #1<1000%
{\hspace*{0pt}#1}%
\else%
{\numprint{#1}}%
\fi%
\fi%
\fi%
}

\newcommand{\ShowPercentage}[2]{%
\FPeval\percentage{round(#1/#2*100,0)}%
\FPeval\percentageOneDecimal{round(#1/#2*100,1)}%
\ifnum \percentage=0%
{\np[\%]{\FPprint{percentageOneDecimal}}}%
\else%
\ifnum \percentage<10%
{\np[\%]{\FPprint{percentageOneDecimal}}}%
\else%
{\np[\%]{\FPprint{percentageOneDecimal}}}%
\fi%
\fi%
\xspace
}
\newlength\BARSIZE  
\newcommand{\inlinechart}[2]{%
\FPeval{\BLACKBARSIZE}{#1/#2}\textcolor{black!80}{\rule{\BLACKBARSIZE\BARSIZE}{1.6ex}}%
\FPeval{\BLACKBARSIZE}{1 - (#1/#2)}\textcolor{black!10}{\rule{\BLACKBARSIZE\BARSIZE}{1.6ex}}%
}

\newcommand*\ChartSmall[3][v]{%
\ifx q#1%
    \np{#2}/\np{#3}(\ShowPercentage{#2}{#3})\else%
\ifx p#1%
    \np{#2}(\ShowPercentage{#2}{#3})\else%
\ifx c#1%
    \inlinechart{#2}{#3}%
\else%
    \np{#2}%
    \ifx r#1%
        /\np{#3}%
    \fi%
    \hspace*{0.5ex}(\ShowPercentage{#2}{#3}) %
    \inlinechart{#2}{#3}%
    \xspace
\fi\fi\fi%
}

\begin{document}

\author{
\IEEEauthorblockN{Nicolas Harrand\IEEEauthorrefmark{1}, Thomas Durieux\IEEEauthorrefmark{1}, David Broman\IEEEauthorrefmark{1}, and Benoit Baudry\IEEEauthorrefmark{1}}

\IEEEauthorblockA{
\IEEEauthorrefmark{1}\textit{EECS and Digital Futures, KTH Royal Institute of Technology, Stockholm, Sweden}\\ Email: \{harrand, tdurieux, dbro, baudry\}@kth.se\\
}
}

\title{The Behavioral Diversity of  Java JSON Libraries}

\maketitle

\def\nbTotalProject{147991}

\newcommand{\nbsrcjson}{$5$\@\xspace}
\newcommand{\nbtargetjson}{$20$\@\xspace}

\newcommand{\nbcorrectjson}{$206$\@\xspace}
\newcommand{\nberroredjson}{$267$\@\xspace}
\newcommand{\nballjson}{$492$\@\xspace}

\newcommand{\orgjson}{\texttt{org.json}\@\xspace}
\newcommand{\jsonsimple}{\texttt{json-simple}\@\xspace}
\newcommand{\gson}{\texttt{gson}\@\xspace}
\newcommand{\fastjson}{\texttt{fastjson}\@\xspace}
\newcommand{\jackson}{\texttt{jackson}\@\xspace}

\newcommand{\wfcorpus}{\texttt{Well-formed}\@\xspace}
\newcommand{\ifcorpus}{\texttt{Ill-formed}\@\xspace}

\newcommand{\yasjf}{\texttt{argo}\@\xspace}
\newcommand{\yasjfapi}{\texttt{argo-api}\@\xspace}



\begin{abstract}

\json is an essential file and data format in domains that span scientific computing, web APIs or configuration management. Its popularity has motivated significant software development effort to build multiple libraries to process \json data. Previous studies focus on performance comparison among these libraries and lack a software engineering perspective.

We present the first systematic analysis and comparison of the input / output behavior of \nbtargetjson  \json libraries, in a single software ecosystem: Java/Maven. We assess behavior diversity by running each library against a curated set of 473 \json files, including both well-formed and ill-formed files. 
The main design differences, which influence the behavior of the libraries, relate to the choice of data structure to represent \json objects  and to the encoding of numbers.  We observe  a remarkable behavioral diversity with ill-formed files, or corner cases such as large numbers or duplicate data. Our unique behavioral assessment of \json libraries paves the way for a robust processing of ill-formed files, through a multi-version architecture.


\end{abstract}



\section{Introduction}

\sloppy
JavaScript Object Notation, or \json, is a ubiquitous file and data exchange format. It is used in domains that span web APIs \cite{tan2016service}, scientific computing \cite{millman2014developing}, data management \cite{LiuHMCLSSSAA20}, or configuration management \cite{wittern2016look}. 
Despite the importance of \json in software applications of all kinds,  very few works analyze the software engineering aspects of the \json ecosystem \cite{BarbagliaMC17,Habib2021}
Previous research works about \json focus on  data representations through schema definitions and inference \cite{pezoa2016foundations,bourhis2017json}, and efficient algorithms for processing  \json files \cite{LangdaleL19,BaaziziCGS19}. 
Meanwhile, the massive adoption of \json has motivated important software development efforts, leading to the release and maintenance of many libraries to process \json files in different languages. 
Our work focuses on a systematic comparison of the input/output behavior of independent \json libraries, in Java.

Behavioral diversity  can be harnessed for improving reliability~\cite{avizienis85,PopovSS12}, performance~\cite{shacham2009chameleon}, or testing~\cite{sondhi2019similarities,boussaa2020leveraging}. 
All these techniques require one foundation: a sound assessment of the existing diversity that can be exploited.  For example, Koopman and DeVale quantified the diversity of failure modes among Posix implementations \cite{koopman1999comparing}  and  Srivastava and colleagues assessed the diversity of vulnerabilities among implementations of the Java Class Library \cite{srivastava2011security}.

Our work proposes the first assessment of the behavior diversity among Java \json libraries.
The focus on one single programming language allows for a precise behavior analysis based on the language's semantics. 
We systematically analyze  \nbtargetjson  \json libraries. This set includes a variety of implementations, from the popular \jackson and \gson libraries, to more uncommon ones such as \texttt{cookjson} or \texttt{sojo}. Our analysis of these libraries consists of two steps. 
First, we determine how each development team decides to represent the $6$ \json types defined in the RFC 8259 standard \cite{rfc8259}.
Second, we analyze the input/output behavior of the libraries.  For this, we curate a novel, diverse set of  473 \json files to be processed by each library.
This set includes \nbcorrectjson well-formed and \nberroredjson ill-formed \json files. We run each library with all the files, and we assess  whether each library has a behavior \conform to the RFC 8259 standard. We analyze the behavior of each separate library, as well as to what extent the libraries collectively behave the same for a portion of the files. The latter analysis is what we call behavior diversity: for what parts of the inputs the libraries behave the same or differently.

We observe significant variations in the representation of \json types. Two of these design decisions have a major impact on the libraries' behavior: the choice of an ordered or unordered data structure to represent the \json objects, and the representation of numbers. The behavioral analysis reveals that the libraries behave globally well when processing the well-formed files: 17 out of \nbtargetjson libraries behave \conform to the standard in more than 80\% of the cases. The behavior variations when processing well-formed files relate to corner cases such as very big numbers or duplicate keys. The corpus of ill-formed \json files reveals a significantly larger behavior diversity. A majority of the libraries exhibit a non \conform behavior for more than $20\%$ of the inputs, and the libraries behave the same only for $0.75\%$ of the input files. Yet, when considering the whole set of libraries, $99.3\%$ of the files are processed correctly by one library at least. This  suggests the opportunity for a resilient multi-version architecture \cite{avizienis85,xu17b} for \json.

The main contributions of this work are as follows
\begin{itemize}[nosep]
    \item The first systematic analysis of the input/output behavior of \nbtargetjson independent Java \json libraries.
    \item An exhaustive cartography of the different representations of \json types in Java.
    \item Empirical evidence of significant behavior diversity among the \json libraries when they process ill-formed files. 
\end{itemize}


\section{Background about \json}
\label{sec:foundations}

This section introduces  the \json format, as well as the \json specifications. We hint at several factors that lay the foundations for diverse implementations.

\subsection{The \json format}

\json is a file and data exchange format. It provides a textual representation of data that is readable by  humans and machines. The \json format proposes $6$ different types: $2$ composite types and 4 primitive ones. Primitive types are \texttt{Strings}, \texttt{Numbers}, \texttt{Booleans} and \texttt{null}. The composite types are \texttt{Object} types that map \texttt{String} keys to values of any type, and an \texttt{Array} type that is an ordered collection of elements of any type. This allows for arbitrary depth of nested data. \autoref{lst:json-example} shows an example of a \json document. It is composed of an object with four keys,
\texttt{"awardYear"} and \texttt{"prizeAmount"} are associated with numbers, \texttt{"category"} is associated with a nested object, and \texttt{"laureates"} with an array.

\begin{lstlisting}[language=json,firstnumber=1, caption={Excerpt from a \json file returned by nobelprize.org when searching for the list of laureates.}, label={lst:json-example}]
{"awardYear": 1901,
 "category":{"en":"Chemistry","no":"Kjemi","se":"Kemi"},
 "prizeAmount": 150782,
 "laureates":[{
      "knownName":{"en":"Jacobus H. van 't Hoff"}}]}
\end{lstlisting}

\subsection{\json specifications}

We introduce three root causes that support diverse behaviors for \json libraries: the evolution of \json specifications over time; the ambiguities and explicit flexibility of the specifications; design and implementation choices.

Four consecutive IETF Requests for Comments (RFCs) specify  the \json format: RFC 4627~\cite{rfc4627} in 2006, RFC 7158~\cite{rfc7158} in 2013, RFC 7159~\cite{rfc7159} in 2014, and RFC 8259~\cite{rfc8259} in 2017. It is important to note that the RFCs were released after  \json  had been used for several years (since the early 2000s), as an attempt to standardize existing usages. In this work, we rely on the most recent specification, RFC 8259~\cite{rfc8259}, to define what constitutes a valid \json text or not. Meanwhile, the development of several of the libraries we study started before the publication of this latest RFC. Some of them have been updated since 2017.

The different RFC versions refine the specification of the  \json format, making it less and less ambiguous.  For example, the first RFC (4627) stipulates that \textit{``octal and hex forms are not allowed''}, implicitly allowing other forms such as decimal and binary. The later RFCs refine this, explicitly stating that numbers are only \textit{``represented in base 10''}. There is one major exception in the refinement process of \json specifications. RFC 7158~\cite{rfc7158} introduces an evolution that makes previously invalid \json documents valid. It stipulates that a \json text \textit{``can be any \json value, removing the constraint that it be an object or array''}. These changes in the specification illustrate how libraries developed at different times may differ in what they consider as valid inputs.

\json RFCs explicitly leave room for library developers to choose what their parser accepts. RFC 8259 states that a \textit{``\json parser MUST accept all texts that conform to the \json grammar''} but a \textit{``\json parser MAY accept non-JSON forms or extensions''}. On the other hand, \json libraries that serialize objects into text \textit{``MUST strictly conform to the \json grammar''} to respect the specification. \textit{``The names within an object SHOULD be unique''.} 

\subsection{Design choices for \json libraries}

In this paper, we analyze the behavior of \json libraries, in Java. These  libraries all expose functions to parse \json files into \json objects, e.g., to process the data or to store it in a database, as well as functions to serialize \json objects into a \json file, e.g., to exchange data between different services. 

Library developers are free to choose how they represent the 6 \json types in Java. The \texttt{Object} \json type is a composite type that includes \texttt{(key,value)} pairs, which can be represented with any kind of Map or an ad-hoc data structure to represent the pairs. The \texttt{Array} \json type is also composite and can be represented with any kind of list or array. The \texttt{String}  \json type can be represented directly with the primitive Java String type, or with ad-hoc type that encapsulates a String. \json \texttt{Number}s can be represented with a String or any of the number types in Java (e.g., \texttt{float}, \texttt{long}, etc.). The \json \texttt{Boolean} type can be mapped on to an \texttt{Enum}, a \texttt{bool} or even a \texttt{String}. \json's \texttt{Null}  can be encoded directly in the Java \texttt{null} or represented with an \texttt{Enum}, or a \texttt{String}.

\section{Experimental Protocol}
\label{sec:methodology}


This section presents our  research questions, the set of \json libraries we study, as well as the corpora of files we use to assess their behavior. Then, we detail the protocol to answer each  question.

\subsection{Research Questions}
\label{sec:rq}

\newcommand{\RQone}{\textbf{RQ1. To what extent do \json libraries implement different design choices?}}

\newcommand{\RQtwo}{\textbf{RQ2. How does each library behave on well-formed input JSON files?}}
\newcommand{\RQthree}{\textbf{RQ3. How does each library behave on ill-formed input JSON files?}}

\newcommand{\RQfour}{\textbf{RQ4. Do the 20 \json libraries collectively behave differently on both well-formed and ill-formed JSON files?}}

\RQone

This question investigates how each library represents the \json types  with Java types. 
These choices represent a source of design diversity that can impact the behavior of the libraries.

\RQtwo

RFC 8259 specifies that \textit{``A \json parser MUST accept all texts that conform to the \json grammar''}~\cite{rfc8259}. 
With RQ2, we observe how each library addresses this point of the specification and handles well-formed \json files. 

\RQthree

RFC 8259 specifies that \textit{``A \json parser MAY accept non-\json forms or extensions''}~\cite{rfc8259}. 
In RQ3, we investigate to which extent the developers of \json libraries take advantage of the specification's ambiguity to process ill-formed files.

\RQfour

This question compares the behavior of our \json libraries and quantifies the diversity of software behavior within this set.  Here, we hypothesize that the diversity of behaviors is larger for ill-formed files, as each implementation has to take independent decisions about how they handle these cases.

\subsection{JSON libraries}
\label{sec:lib}


To build our collection of independent Java libraries that implement \json processing capabilities, we start from \href{https://json.org}{https://json.org}. This is the official \json website, setup by the lead author of \json, Douglas Crockford. We visited the site on November 2020, we found a list of $22$  Java libraries. Our goal is to systematically analyze all these libraries, that have been curated by a third-party, authoritative computer scientist. We ignore $3$ of them (\texttt{StringTree}, \texttt{Json-taglib}, \texttt{Fossnova-json}), which we cannot build. 
We compare this dataset to Maven Central ~\cite{jackson-pop}, which naturally hosts a diversity of \json libraries \cite{soto2019emergence}. We find that only \jackson is among the 20 most popular Java \json libraries and is not in our dataset. Consequently, we add \jackson to the dataset for our study.
Our set of libraries includes the latest version of each library, available on Maven Central on November 24th 2020. This constitutes our dataset of \nbtargetjson \json libraries. 

\begin{table}[t]
\centering
\rowcolors{2}{gray!25}{white}
\setlength{\tabcolsep}{2pt}
\begin{tabular}{lrrrr}
  \toprule
\textsc{Library} & \textsc{\# Commits} & \textsc{\# Stars} & \textsc{Version} & \textsc{Last activity} \\ 
  \midrule
  cookjson & 116 & 3 & 1.0.2 & Sept 2017 \\ 
  corn & - & - & 1.0.8 & Feb 2014 \\ 
  fastjson & 3793 & 1.4k & 1.2.75 & Nov 2020 \\ 
  flexjson & - & - & 3.3 & Oct 2014 \\ 
  genson & 395 & 193 & 1.6 & Dec 2019 \\ 
  gson & 1485 & 18.8k & 2.8.5 & May 2020 \\ 
  jackson & 7382 & 2.7k & 2.12.0-rc2 & Nov 2020 \\ 
  jjson & 216 & 12 & 0.1.7 & Jul 2016 \\ 
  johnzon & 780 & - & 1.1.8 & Nov 2020 \\ 
  json & 841 & 3.7k & 20201115 & Nov 2020 \\ 
  json-argo & - & - & 5.13 & Nov 2020 \\ 
  json-io & 1040 & 268 & 4.12.0 & Oct 2013 \\ 
  json-lib & - & - & 3.0.1 & Dec 2010 \\ 
  json-simple & 30 & 594 & 1.1.1 & Jul 2014 \\ 
  json-util & 464 & 48 & 1.10.4-java7 & Oct 2016 \\ 
  jsonij & 348 & - & 0.3.1 & Feb 2020 \\ 
  jsonp & 530 & 75 & 2.0.0 & Nov 2018 \\ 
  mjson & 79 & 67 & 1.4.0 & May 2019 \\ 
  progbase & - & - & 0.4.0 & Nov 2019 \\ 
  sojo & - & - & 1.0.13 & Feb 2019 \\ 
   \bottomrule 
\end{tabular}
    \vspace{0.5em}
    \caption{Description of the \nbtargetjson Java \json libraries under study.\\ A '-' indicates a library that has no public code repository.}
    \label{tab:libs}
    \vspace{-1em}
\end{table}

\autoref{tab:libs} describes the libraries in our dataset. The libraries are presented in alphabetical order. When a version control system is available, we collect the number of commits. When the library is associated with a GitHub repository, we note the number of stars. Column \textsc{Version} contains the latest version available in Maven Central on November 24th 2020.  Column \textsc{Last Activity} gives the date of the last commit if available, or the publication date of the artifact on Maven Central otherwise.

\subsection{\json corpora}
\label{sec:corpora}

To assess the behavior diversity of \json libraries, we execute all of them against  a collection of  \nballjson \json files. We aggregate $4$ \json corpora that were previously  assembled to  benchmark \json libraries.  By doing so, we gather \json files from diverse sources in order to offer a broad coverage of the \json format.: 
\begin{itemize}[nosep]
    \item The official test suite of \url{json.org}~\cite{jsonChecker} that is meant to evaluate the compliance of a \json parser to the \json grammar. It includes 36 \json files labelled as \texttt{pass} or \texttt{fail}.
    \item The Native \json Benchmark~\cite{nativejson} is used to evaluate the performance of native C/C++ \json libraries as well as compliance to RFC 7159. 
    \item The suite  used for "Parsing \json is a minefield"~\cite{jsonTestSuiteMinfield} a study of the challenges and corner cases that a developer may encounter while implementing a \json library. It includes 318 \json files labeled as \texttt{y} for yes, \texttt{n} for no, and \texttt{i} for syntactically correct files that RFC 8259 mentions as potentially problematic.
    \item The test suite of jansson~\cite{jansson} that includes 114 \json files labeled as \texttt{invalid} or \texttt{valid}. We include this test suite because jansson is a popular, open source library, from a different ecosystem (C language). 

\end{itemize}

\begin{table}[t]
\centering
\rowcolors{2}{gray!25}{white}
\begin{tabular}{lrrr}
  \toprule
\textsc{source} &  \textsc{\# Well-formed} & \textsc{\# Ill-formed} & \textsc{Size} \\ 
  \midrule
json.org & 3 & 33 & 2.7kB\\
Native \json & 33 & 33 & 4.6MB\\
minefield & 130 & 188 & 354kB\\
jansson & 46 & 68 & 101.8kB\\
  \midrule
  \midrule
Corpora & \nbcorrectjson & \nberroredjson & 5.1MB\\
   \bottomrule 
\end{tabular}
    \vspace{0.5em}
    \caption{Description of the corpora of \json files used as input to assess the diversity of behavior of \json libraries}
    \label{tab:corpora}
    \vspace{-2em}
\end{table}

We collect a total of $534$ \json files. We remove $42$ duplicated files, as well as $19$ files that can not be read by Java CharsetDecoder class with UTF-8 or UTF-16 encoding. 
\autoref{tab:corpora} summarizes the origin of the \json files as well as the content of our corpora (last line of the table). 
These files, of various sizes, include all types of \json data, including nested data, large Strings and Numbers, or simply very large files, e.g., the Native \json test suite  includes $3$ files bigger than 1MB. The distinction between \wfcorpus and \ifcorpus files is based on the classification established by the authors of the original datasets.
The well-formed \json corpus includes \nbcorrectjson files that are syntactically correct, according to the \json grammar specified in RFC 8259~\cite{rfc8259}. The ill-formed \json corpus includes \nberroredjson files that include some structural errors. These corpora are available online~\cite{reproCorpora}.


\subsection{Protocol for RQ1}

RQ1 explores how the developers of the libraries implement the standard \json types in Java. 
The RFC 8259 \json standard~\cite{rfc8259} describes $6$ types for \json: \texttt{Object}, \texttt{Array}, \texttt{String}, \texttt{Number}, \texttt{Boolean} and \texttt{Null}. By contrast, the Java standard library provides many types that can be used to represent \json types.

To identify the design decisions implemented in each library, we manually explore their source code to find the classes that represent \json data. In particular, we note if Java types are used directly to represent \json values, are extended, or wrapped in a class provided by the library. 

To analyze the implementation of \json Numbers, we execute the libraries to parse \json numeric values that correspond to extreme values of Java types (for instance, values $-2147483648$ and $2147483647$ for 32-bit integers (\texttt{int}) and values $4.9E-324$, $2.2250738585072014E\!-\!308$ and $1.7976931348623157E308$ for 64-bit floating-point numbers (\texttt{double})). We collect the Java objects that are created at runtime.
This allows us to determine the diverse Java types used by the library to represent \json Numbers. The complete list of tested values is available in the reproduction package~\cite{reproTypes}.

\subsection{Protocol for RQ2}
\label{sec:data-json-file-c}
\begin{algorithm}[t]
\scriptsize
  \SetAlgoLined
  \SetNoFillComment

  \SetKwInput{KwData}{Inputs}
  \SetKwInput{KwResult}{Result}
  \SetKwProg{try}{try}{:}{}
  \SetKwProg{catch}{catch}{:}{end}
  \KwData{\\- $\mathit{jsonInput}$: A well-formed \json file, \\- $\mathit{library}$: A \json library} 
  \KwResult{[\conform, \silent, \error]}
    \try{}{
        $\mathit{jsonObject} \gets \underline{parse}(\mathit{library},\mathit{jsonInput})$
        
        \If{$\mathit{jsonObject} = NULL \land \mathit{jsonInput} \neq ``null"$}{
            $\mathit{log}(\mathit{Null\_Object})$
            
            \Return{\texttt{Error}}
        }
    }
    \catch{Exception}{
        \eIf{$\mathit{isChecked}(\mathit{Exception})$}{
            $\mathit{log}(\mathit{Parse\_Exception})$
        }{
            $\mathit{log}(\mathit{Crash})$
        }
        \Return{\texttt{Error}}
    }
    \try{}{
        $\mathit{jsonOut} \gets \underline{serialize}(\mathit{library},\mathit{jsonObject})$
        
        \If{$\mathit{jsonOut} = \mathit{jsonInput}$}{
             $\mathit{log}(\mathit{Equal})$
             
            \Return{\conform}
        }
        \eIf{$\mathit{jsonObject} \equiv \underline{parse}(\mathit{library},\mathit{jsonOut})$}{
             $\mathit{log}(\mathit{Equivalent\_Object})$
             
            \Return{\conform}
        }{
           $\mathit{log}(\mathit{Non\_Equivalent\_Object})$
           
            \Return{\texttt{Silent}}
        }
    }
    \catch{Exception}{
        \eIf{$\mathit{isChecked}(\mathit{Exception})$}{
            $\mathit{log}(\mathit{Print\_Exception})$
        }{
            $\mathit{log}(\mathit{Crash})$
        }
        \Return{\texttt{Error}}
    }
\caption{Test sequence to assess the behavior of \json libraries with \wfcorpus \json files.}
\label{alg:meta-algo-wf}
\end{algorithm}

RQ2 assesses the behavior of each of the \nbtargetjson \json libraries in our dataset on \wfcorpus \json files. 
The protocol consists in executing each library, passing every file in the \wfcorpus corpus as input.
We categorize the outcome of each execution as \conform to the standard, \texttt{Error} or \texttt{Silent}. We consider a library as \conform when it correctly parses and serializes a \json file that is \wfcorpus according to RFC 8259 \cite{rfc8259}.
An \texttt{Error} behavior is when the library explicitly notifies an issue, e.g., with an exception, while a \texttt{Silent} behavior indicates that  the library does not explicitly notify  an issue.



Algorithm \ref{alg:meta-algo-wf} defines the sequence of operations we execute with each library. It takes a \wfcorpus JSON file and a library as input, and returns one of the alternatives \conform, \texttt{Error}, and \texttt{Silent}.
The library parses a \json file into a \json object (step 1, line 2), then it serializes the object back into a file (step 2, line 13). If this file is strictly equal to the input file, the behavior is \conform, otherwise, we parse the second file back into a \json object (step 3, line 17). If the two objects produced after both parsing are equivalent, the behavior is \conform to the specification. Any other exceptional behavior crash is  an  \error. If the objects at step 1 and step 3 are not equivalent and the library does not notify it, this is a \silent behavior. 

For each execution we log intermediate behavior: Equal (EQ) when the input file and the file produced at step 2 are strictly equal (not case-sensitive) (Line 14);  
 Equivalent\_Object (EV) when the  Java objects retrieved at step 1 and at step 3 are equivalent;  Non\_Equivalent (NE) when the two objects are not equivalent (Line 21); Null\_Object (NO)  in step 1 (Line 4), when parsing produces a \texttt{null} object, which is not a representation of the a \texttt{null} \json value. The execution of \autoref{alg:meta-algo-wf} can be interrupted by exceptions. The algorithm distinguishes between checked exceptions, that have been anticipated by the developers, from unchecked exceptions that lead to a crash. We observe 2 types of checked exceptions, in Line 8 (Parse\_Exception (PA) in step 1),  and in line 25 (Print\_Exception (PR)  in step 2). A  Crash (CR) can occur in lines 10 and 27.

\begin{algorithm}[t]
\scriptsize
  \SetAlgoLined
  \SetNoFillComment
  
  \SetKwInput{KwData}{Inputs}
  \SetKwInput{KwResult}{Result}
  \SetKwProg{try}{try}{:}{}
  \SetKwProg{catch}{catch}{:}{end}
  \KwData{\\- $\mathit{jsonInput}$: An ill-formed \json file, \\- $\mathit{library}$: A \json library} 
  \KwResult{[\conform, \silent, \error]}
  
    \try{}{
        $\mathit{jsonObject} \gets \underline{parse}(\mathit{library},\mathit{jsonInput})$
        
        \eIf{$\mathit{jsonObject} = NULL$}{
            $\mathit{log}(\mathit{Null\_Object})$
            
            \Return{\conform}
        }{
            $\mathit{log}(\mathit{Unexpected\_Object})$
            
            \Return{\silent}
        }
    }
    \catch{Exception}{
        \eIf{$\mathit{isChecked}(\mathit{Exception})$}{
            $\mathit{log}(\mathit{Parse\_Exception})$
            
            \Return{\conform}
        }{
            $\mathit{log}(\mathit{Crash})$
            
            \Return{\error}
        }
    }

\caption{Test sequence to assess the behavior of \json libraries with \ifcorpus \json files.}
\label{alg:meta-algo-if}
\end{algorithm}

At step 3 on line 17, we check the equivalence between two objects according to the following rules: \json arrays contain only equivalent elements in the same order, \json objects include the same set of keys, and for each key, an equivalent object, strings are strictly equal, numbers are equal and of the same type, literals  are equal.

\subsection{Protocol for RQ3}
\label{sec:data-json-file-e}

RQ3 assesses the behavior of each  library with \ifcorpus \json files. 
Each library tries to parse each \ifcorpus file. This operation can result in 3 different behaviors, as described in \autoref{alg:meta-algo-if}. 
The library is \conform to the standard if it recognizes the input file as \ifcorpus and explicitly notifies so. This manifests as a Null\_Object (NO) or a fail with an explicit Parse\_Exception (PA). 
A library behaves \silent if it accepts to parse the \ifcorpus file and generates an Unexpected\_Object (UO), without an explicit notification (line 7). 
If the library crashes (CA), we classify this as an \texttt{Error} behavior.

\subsection{Protocol for RQ4}
\label{sec:prot3}
In this research question, we investigate for which  \json files the libraries behave the same or have diverse behaviors. We make the hypothesis that the diversity of behaviors among \json Java libraries is greater when processing  \ifcorpus inputs rather than \wfcorpus inputs. 

First, we assess the behavioral diversity pairs of libraries with the behavioral distance defined in  \autoref{metric:distance}.

\begin{metric}
\label{metric:distance} \textbf{\emph{Behavioral distance}}. We adapt Jaccard's distance to determine the behavioral diversity between two libraries that execute with the same set of input files.  
Given  $C$, a corpus of input files, two libraries $l_1$ and $l_2$, the behavioral distance $bd_{C}(l_{1},l_{2})$ between the two libraries is the probability that the two libraries behave differently on an input file picked in  C:
\begin{equation*}
    bd_{C}(l_{1},l_{2}) = \frac{|\{f \in C | outcome_{l_{1}}(f) \neq outcome_{l_{2}}(f)\}|}{|C|}
\end{equation*}

\end{metric}

Second, we assess the global diversity among all \nbtargetjson libraries. For this, we analyze the proportion of files for which a part of the libraries behaves the same. 

 



\section{Results}
\label{sec:results}
In this section, we describe and discuss the findings after performing the experiments according to the protocols and research questions described in Section~\ref{sec:methodology}.

\subsection{\RQone}
\label{sec:rq1}

\begin{table*}[ht]
\centering
\rowcolors{2}{gray!25}{white}
\begin{tabularx}{\textwidth}{X|m{15em} m{7em} l m{13em} lll}
  \toprule
  \textsc{Library} & \textsc{Object • Key} & \textsc{Array} & \textsc{String} & \textsc{Number} & \textsc{Boolean} & \textsc{Null}  \\ 
  \midrule
  cookjson & \textit{E} HashMap • String & \textit{E} ArrayList & \textit{C} String & \textit{C} BigDecimal, Long, Double, Integer, byte[] & Enum & Enum  \\ 
  corn & \textit{C} ConcurrentHashMap • String & \textit{C} CopyOn-WriteArrayList & \textit{C} String & \textit{C} String & \textit{C} String & \textit{T}  \\ 
  fastjson & C HashMap/LinkedHashMap • String & C ArrayList & String & Integer, Long, BigInteger, BigDecimal & Boolean & null  \\ 
  flexjson & HashMap • String & ArrayList & String & Long, Double & Boolean & null  \\ 
  genson & HashMap • String & ArrayList & String & Long, Double & Boolean & null  \\ 
  gson & \textit{C} LinkedTreeMap • String & \textit{C} ArrayList & \textit{C} Object & \textit{C} Object & \textit{C} Object & \textit{T}  \\ 
  jackson & \textit{C} LinkedHashMap • String & \textit{C} ArrayList & \textit{C} String & \textit{C} int, long, double, float, short, BigDecimal, BigInteger & \textit{C} boolean & \textit{T}  \\ 
  jjson & \textit{C} HashMap • String & \textit{C} ArrayList & \textit{C} StringBuffer & \textit{C} String & \textit{C} boolean & T  \\ 
  johnzon & C UnmodifiableMap \textit{E} AbstractMap • String & \textit{C} List \textit{E} AbstractList & \textit{C} String & \textit{C} BigDecimal, double, long & \textit{C} Enum & \textit{C} Enum  \\ 
  json & \textit{C} HashMap • String & \textit{C} ArrayList & String & Integer, BigDecimal & Boolean & \textit{T}  \\ 
  json-argo & \textit{C} LinkedHashMap • JsonStringNode & \textit{C} List & \textit{C} String & \textit{C} String & \textit{C} Enum & \textit{C} Enum  \\ 
  json-io & \textit{E} LinkedHashMap • Object & Object[] & String & Long, Double & Boolean & null  \\ 
  json-lib & \textit{C} ListOrderedMap • String & \textit{C} ArrayList & String & Double, Integer & Boolean & \textit{T}  \\ 
  json-util & LinkedHashMap • String & ArrayList & String & Long, Double & Boolean & null  \\ 
  jsonij & \textit{C} LinkedHashMap • String & \textit{C} ArrayList & \textit{C} String & \textit{C} double, long, Number & Enum & Enum  \\ 
  jsonp & \textit{C} UnmodifiableMap \textit{E} AbstractMap • String & \textit{C} List & \textit{C} String & \textit{C} int, long, BigDecimal & \textit{C} Enum & \textit{C} Enum  \\ 
  json-simple & \textit{E} HashMap • Object & \textit{E} ArrayList & String & Long, Double & Boolean & null  \\ 
  mjson & \textit{C} HashMap • String & \textit{C} ArrayList & \textit{C} String & \textit{C} Number & \textit{C} boolean & \textit{T}  \\ 
  progbase & HashMap • String & ArrayList & String & Double, Integer & Boolean & null  \\ 
  sojo & LinkedHashMap • String & ArrayList & String & Long, Double & Boolean & null  \\ \hline\hline
  Alternatives & 13 & 7 & 4 & 12 & 6 & 4   \\ 
  \bottomrule
\end{tabularx}
    \vspace{0.5em}
\caption{A cartography of the different representations of \json types in Java \json libraries. We mark with a \textit{C} class that contains another type and delegates calls to it. We mark with \textit{E} a type that extends another. We mark with a \textit{T} the cases where a library defines a new type that does not delegate calls to any type of the standard library.}
    \vspace{-1em}
\label{tab:design}
\end{table*}

In this research question, we study how the \nbtargetjson \json libraries of this study map the different \json type to Java types. 
\autoref{tab:design} summarizes these design decisions. 
Each column in the table represents one  \json type, and their content  indicates how developers have chosen to represent them. We mark \texttt{C Type} (Contains) the cases when a library defines its own type that delegates calls to the after mentioned the standard Java \texttt{Type}. For example, the  \texttt{corn} library represents \json Objects with the class \texttt{net.sf.corn.converter.json.JsTypeComplex}. This class stores the  key/values pairs in a standard Java \texttt{ConcurrentHashMap}, the keys are stored as Strings. We mark \texttt{E} (Extends)  the cases when a library defines a type that inherits directly from the after-mentioned Java type. 
For example, \texttt{cookjson} represents \json Objects with the class \texttt{org.yuanheng.cookjson.value.CookJsonObject} that directly extends the class \texttt{HashMap} from the standard library.
We mark \texttt{T} the cases when a library defines its own type to represent a \json type and does not rely on any standard Java type.  For example, the  \texttt{corn} library represents \json Null with the class \texttt{net.sf.corn.converter.json.JsTypeNull}. The last line in \autoref{tab:design} indicates the number of different ways to represent a \json type among our set of libraries.

\json Objects are always represented, one way or another, with the Map interface from the Java standard library. Yet, the implementation and wrapping vary among libraries.
We observe that $10$ libraries use an ordered map to store \json objects key-value pairs, while $7$ use a \texttt{HashMap} that does not preserve the order of insertion. \fastjson uses either one of these data structures, depending on an option. Both \texttt{JsonP} and \texttt{Johnzon} collections' are non-modifiable. The decision of using an ordered map directly impacts whether or not the operation of parsing and serialization leaves a \json text unchanged syntactically. 

\json arrays are mapped to the List interface, except for \texttt{json-io} that relies on primitive \texttt{Object} array. $15$ out of \nbtargetjson use an \texttt{ArrayList} either directly, by extending the class, or by wrapping it in a container class. \texttt{corn} is the only library to use a \texttt{CopyOnWriteArrayList}.

\json numbers are mapped to many different types depending on the library. For example, \texttt{sojo} relies on \texttt{Long} for integers and \texttt{Double} for real numbers. This means that the library cannot represent numbers that are larger than $2^{63}$, or values more precise than $2^{-1022}$, since those types from the standard library use $64$ bits representation. Some libraries use primitive types, e.g., \jackson, or their boxed version, e.g., \texttt{flexjson}. \texttt{corn} even stores a textual representation of the \json numbers in a \texttt{String} and lets its clients decide which numeric type to use.

\json Booleans are represented by the Java Boolean class in $10$ libraries. $5$ libraries define an enum to represent all three \json literals (\texttt{TRUE}, \texttt{FALSE} and \texttt{NULL}).
$8$ libraries represent the \json value \texttt{NULL} with a \texttt{null} Java object. All others represent it with a specific class, or enum. Note that a library that does use the Java \texttt{null} value to represent the \json literal \texttt{NULL}, cannot use it as a mechanism to communicate a missing key to the library's client.


We observe that the libraries that define their own types, still heavily reuse standard types (i.e., very few cases of \texttt{T}). The most popular way of reusing standard types is through containment and delegation ($54$ occurrences among the $120$ choices analyzed in \autoref{tab:design}). We also observe that some libraries do extend directly types from the standard library instead. This design decision has an impact on the API that the \json library exposes: a class that encapsulates a standard type exposes its own public interface, while a class that inherits a standard class also exposes the inherited API.

Overall, there is not a single \json type that is universally mapped to the same Java type by all \nbtargetjson libraries.
The number of  classes implemented in the different  libraries to represent \json types varies from $0$ to $13$. On one extreme, libraries such as \texttt{flexjson}, \texttt{genson}, \texttt{json-util}, \texttt{progbase} and \texttt{sojo} do not implement any specific class to model \json types. Their parsers directly return Java objects from types provided by the standard library. Their \json generator, also directly accepts Java objects and serializes them to \json text. On the other end of the spectrum, libraries such as \texttt{cookjson}, \texttt{johnzon} and \texttt{jsonp} or \jsonsimple implement specific classes for \json Objects and Arrays, but represent \json Strings, Numbers and Booleans as Java boxed type.

The last line of \autoref{tab:design} emphasizes this wide diversity of design choices. The \nbtargetjson libraries exhibit up to 13 different choices to represent Objects and 12 different choices to represent Numbers. Even the choice of String representations, which can be trivial with the standard \texttt{java.lang.String}, is subject to different choices.
This diversity of design choices does impact the behavior of the libraries. In particular, the choice of whether to use an ordered collection or not, as well as the choice of types used to represent \json numbers directly affects how the library behaves.



\vspace{1em}
\begin{mdframed}[style=mpdframe]\textbf{Answer to RQ1.} The diversity of design decisions among \nbtargetjson libraries is remarkable, with up to 13 different ways of representing \json Objects and 12 ways of representing \json. We note that the choice of an ordered map for objects and the representation of numbers are two key choices that impact the behavior of the libraries, providing diverse trade-offs between performance and usability.
\end{mdframed}

\subsection{\RQtwo}

\begin{table}[th]
\rowcolors{2}{gray!25}{white}
\setlength{\tabcolsep}{1.1pt}
\centering
\begin{tabular}{@{}l|ccc|c|ccccc@{}}
 Library & \multicolumn{3}{c|}{\conform} & \silent & \multicolumn{5}{c}{\error} \\ 
 & EQ & EV & Tot & NE & NO & PA & PR & CR & Tot \\ 
  \midrule
  cookjson &  97 &  91 & 188 (91.3\%) & 0 (0\%) & - & 18 & - & - & 18 (8.7\%) \\ 
  corn & 154 &  51 & 205 (99.5\%) & 0 (0\%) & - & 1 & - & - & 1 (0.5\%) \\ 
  fastjson &  96 &  93 & 189 (91.7\%) & 4 (1.9\%) & - & 9 & 4 & - & 13 (6.3\%) \\ 
  flex-json &  89 & 101 & 190 (92.2\%) & 0 (0\%) & - & 11 & 5 & - & 16 (7.8\%) \\ 
  genson &  87 & 104 & 191 (92.7\%) & 7 (3.4\%) & 8 & - & - & - & 8 (3.9\%) \\ 
  gson & 129 &  77 & 206 (100\%) & 0 (0\%) & - & - & - & - & 0 (0\%) \\ 
  jackson &  98 &  94 & 192 (93.2\%) & 5 (2.4\%) & - & 9 & - & - & 9 (4.4\%) \\ 
  jjson & 125 &  66 & 191 (92.7\%) & 3 (1.5\%) & 9 & - & - & 2 & 11 (5.3\%) \\ 
  johnzon &  92 &  96 & 188 (91.3\%) & 0 (0\%) & - & 18 & - & - & 18 (8.7\%) \\ 
  json & 112 &  86 & 198 (96.1\%) & 3 (1.5\%) & - & 5 & - & - & 5 (2.4\%) \\ 
  json-argo & 135 &  62 & 197 (95.6\%) & 0 (0\%) & - & 9 & - & - & 9 (4.4\%) \\ 
  json-io &  88 &  80 & 168 (81.6\%) & 18 (8.7\%) & - & 15 & 5 & - & 20 (9.7\%) \\ 
  json-lib &  95 &  92 & 187 (90.8\%) & 0 (0\%) & - & 18 & - & 1 & 19 (9.2\%) \\ 
  json-simple &  88 &  98 & 186 (90.3\%) & 0 (0\%) & - & 15 & 5 & - & 20 (9.7\%) \\ 
  jsonij &  96 &  64 & 160 (77.7\%) & 0 (0\%) & - & 41 & 5 & - & 46 (22.3\%) \\ 
  jsonp & 106 &  90 & 196 (95.1\%) & 0 (0\%) & - & 10 & - & - & 10 (4.9\%) \\ 
  jsonutil &  96 &  55 & 151 (73.3\%) & 14 (6.8\%) & - & 34 & 7 & - & 41 (19.9\%) \\ 
  mjson &  91 &  99 & 190 (92.2\%) & 0 (0\%) & - & 11 & 5 & - & 16 (7.8\%) \\ 
  progbase &  76 &  90 & 166 (80.6\%) & 23 (11.2\%) & 9 & - & 2 & 6 & 17 (8.3\%) \\ 
  sojo &  86 &  76 & 162 (78.6\%) & 0 (0\%) & - & 42 & 2 & - & 44 (21.4\%) \\ 
  \midrule
  Population & 184 & 152 & 206 (100\%) & 52 (25.2\%) & 9 & 74 & 17 & 9 & 89 (43.2\%) \\ 
   \bottomrule 
\end{tabular}
    \vspace{0.5em}
    \caption{Observed behavior when running the \json libraries under study with \nbcorrectjson \wfcorpus   files.}
    \label{tab:bench-correct}
    \vspace{-2em}
\end{table}

In this research question, we investigate how the different \json libraries behave when processing \wfcorpus \json files. We apply the protocol described in \autoref{sec:data-json-file-c}.


\autoref{tab:bench-correct} provides an overview of the outcomes on the \wfcorpus corpus. 
The first column gives the name of a library.
The second column provides the number of files for which a library behavior is \conform: number of \texttt{EQUAL} (EQ) outcomes,  \texttt{EQUIVALENT\_OBJECT} (EV) outcomes and the total number of \conform behavior (Tot).
The third column shows the number of times a library behaves as \texttt{Silent}, i.e. NON\_EQUIVALENT (NE) in the case of the well-formed files. 
The fourth column gives the number of \texttt{Error} cases: \texttt{NULL\_OBJECT}(NO), \texttt{PARSE\_EXCEPTION} (PA), \texttt{PRINT\_EXCEPTION} (PR), \texttt{CRASH} (CR) and total number of \texttt{Error} (Tot).
The last line of \autoref{tab:bench-correct} (Population) aggregates results over the whole set of libraries: each column is the number of files in the corpus for which at least one library produces a given outcome.
These aggregate observations indicate how the set of libraries behaves as a whole, with respect to well-formed files.

For example, the first row shows that the behavior of  \texttt{cookjson} is \conform (Tot) for $188$ files ($91.3\%$): $97$ \texttt{EQUAL} (EQ) and $91$ \texttt{EQUIVALENT\_OBJECT} (EV). It also produces an \texttt{Error} behavior for $18$ files ($8.7\%$), where it triggers a \texttt{PARSE\_EXCEPTION} (PA).

The libraries are \conform for at least $73.3\%$ of the \nbcorrectjson \wfcorpus \json files and up to $100\%$ in the case of \gson. 
The libraries exhibit an  \texttt{Error} for a number of \wfcorpus files ranging  from $0\%$ for \gson, up to $22.6\%$ for \texttt{jjson}. We observe that $89$ files ($43.2\%$) trigger an \texttt{Error} for at least one library. 
The share of files leading to \texttt{Silent} outcomes ranges from $0\%$ for $12$ libraries, and up to $11.2\%$ for \texttt{progbase}. There are $59$ files that trigger a \texttt{Silent} outcome for at least one library. 
These results illustrate that all libraries' behaviors are \conform for the vast majority of the \wfcorpus files. Meanwhile, most libraries exhibit  \texttt{Silent} or \texttt{Error} behavior on a few input files. In the following, we discuss the causes of non-\conform behavior.

\begin{lstlisting}[language=json, caption={Syntactically correct \json. Each line is extracted from a file in the well-formed corpus.}, label={lst:json-correct}]
[-0]
[1E22]
[1e+2]
["\u2064"]
{"a":null}
\end{lstlisting}


All libraries produce a large amount of \texttt{EQUIVALENT} outcomes, rather than only \texttt{EQUAL} ones. We find two key reasons for this. First, the standard grammar specification allows several distinct representations for the same value. 
The examples  in \autoref{lst:json-correct} illustrate this phenomenon. \verb|0| and \verb|-0| are two different character sequences that represent the same value; \verb|1E22|, \verb|1e22| and \verb|1e+22| represent the same number. The developers of a \json library need to decide in which way they serialize such values. However, whatever decision they take, it implies that some \json input cannot be serialized back in the same character string. Second, \texttt{Equivalent} outcomes are  related to the choice of an ordered or unordered map to represent \json objects. Both design decisions  comply with the standard specification. Yet, the significant diversity of choices we observed in RQ1 manifests into a diversity of behaviors, as shown among the \conform results of \autoref{tab:bench-correct}.

Library developers sometimes deliberately decide to not strictly follow the specification.
For example, the library \fastjson does not serialize Map entries for which the value is null, if the user does not select the option \texttt{SerializerFeature}.
The example in line 5 of \autoref{lst:json-correct}  is parsed correctly by \fastjson and is serialized into an empty object, which we classify as a \texttt{Silent} behavior. This behavior is intended by the developers of the library.

\begin{lstlisting}[language=json, caption={Syntactically correct \json that the RFC warns against. Each line is extracted from a file of the Well-formed corpus.}, label={lst:json-undefined}]
[9223372036854775808]
[0.4e00669999999999999999999999999999999999999999999999999999999999999999999999999999999999999999999999999999999999999999999969999999006]
{"a":"b","a":"b"}
\end{lstlisting}

The \wfcorpus corpus, is built according to the RFC 8259~\cite{rfc8259}, which allows for a \json document to be a lonely value, i.e., a value that is neither a \json object nor a \json array (for instance a number). RFC 8259 indicates: \textit{``Note that certain previous specifications of \json constrained a \json text to be an object or an array''} \cite{rfc8259}. Indeed, \texttt{cookjson} does trigger a \texttt{PARSE\_EXCEPTION} for the $8$ files that contain lonely values in the corpus.

The \wfcorpus corpus contains only \json files that are syntactically correct. Still, RFC 8259 mentions well-formed cases that may lead to interoperability issues among different parsers. The two paragraphs below discuss the specific cases of numbers and duplicated keys, which lead to \texttt{Error} and \texttt{Silent} outcomes.
\looseness=-1

 \textit{``}[RFC 8259] \textit{allows implementations to set limits on the range and precision of numbers accepted''}~\cite{rfc8259}. 
\autoref{lst:json-undefined} illustrates such cases. For example, line 1 shows an array that contains an integer too big to be expressed with 64 bits. In Java, the \texttt{BigInteger} class can represent this value, and all libraries produce an \texttt{EQUAL} outcome for this case. Line 2 contains a value that cannot be represented by a 64 bit floating-point number, hence only $6$ libraries do not throw any error.
For example, \gson represents the value with \texttt{LazyParsedNumber}. Note that this allows \gson to correctly re-serialize  the value, but a client of the library would still need to process a String representing the value on its own. Libraries that represent numbers with the Java types \texttt{double} and \texttt{long}, as seen in RQ1, are bound to produce either \texttt{Silent} or \texttt{Error} outcomes for values that cannot be represented with these types.

RFC 8259 indicates: 
\textit{``When the names within an object are not unique, the behavior of software that receives such an object is unpredictable''} \cite{rfc8259}. Line 3 of \autoref{lst:json-undefined} illustrates  a case with a duplicate key and a duplicate value. When both a key and its value are duplicated, all libraries except  \orgjson, \texttt{flex-json} and \texttt{jsonij} generate an \texttt{EQUIVALENT} outcome (the duplicated key disappears in the re-serialized version). 

\vspace{1em}
\begin{mdframed}[style=mpdframe]\textbf{Answer to RQ2.} 
\gson is the singular case that is \conform for $100\%$ of the \wfcorpus \json files. 
All other libraries behave globally well, including 16 libraries that correctly process more than $80\%$ of the files and which behave \texttt{Error} with different inputs.  
\end{mdframed}

\subsection{\RQthree}
In this research question, we investigate how the different \json libraries behave when processing \ifcorpus \json files. We apply the protocol described in \autoref{sec:data-json-file-e}. 

\begin{table}[t]
\rowcolors{2}{gray!25}{white}
\setlength{\tabcolsep}{3pt}
\centering
\begin{tabular}{@{}l|ccr|r|r@{}}
  Library & \multicolumn{3}{c|}{\conform} & \silent & \error \\ 
    & PA & NE & Tot & UO & CR \\ 
  \midrule
cookjson & 232 & - & 232 (86.9\%) & 35 (13.1\%) & 0 (0\%) \\ 
  corn & 100 & - & 100 (37.5\%) & 164 (61.4\%) & 3 (1.1\%) \\ 
  fastjson & 179 & 2 & 181 (67.8\%) & 86 (32.2\%) & 0 (0\%) \\ 
  flex-json & 122 & - & 122 (45.7\%) & 142 (53.2\%) & 3 (1.1\%) \\ 
  genson & 160 & 20 & 180 (67.4\%) & 84 (31.5\%) & 3 (1.1\%) \\ 
  gson & 131 & - & 131 (49.1\%) & 136 (50.9\%) & 0 (0\%) \\ 
  jackson & 232 & - & 232 (86.9\%) & 32 (12\%) & 3 (1.1\%) \\ 
  jjson & 12 & 24 & 36 (13.5\%) & 173 (64.8\%) & 58 (21.7\%) \\ 
  johnzon & 235 & - & 235 (88\%) & 29 (10.9\%) & 3 (1.1\%) \\ 
  json & 108 & - & 108 (40.4\%) & 156 (58.4\%) & 3 (1.1\%) \\ 
  json-argo & 247 & - & 247 (92.5\%) & 20 (7.5\%) & 0 (0\%) \\ 
  json-io & 203 & 2 & 205 (76.8\%) & 59 (22.1\%) & 3 (1.1\%) \\ 
  json-lib & 156 & - & 156 (58.4\%) & 111 (41.6\%) & 0 (0\%) \\ 
  json-simple & 202 & - & 202 (75.7\%) & 65 (24.3\%) & 0 (0\%) \\ 
  jsonij & 238 & - & 238 (89.1\%) & 26 (9.7\%) & 3 (1.1\%) \\ 
  jsonp & 231 & - & 231 (86.5\%) & 33 (12.4\%) & 3 (1.1\%) \\ 
  jsonutil & 168 & 2 & 170 (63.7\%) & 94 (35.2\%) & 3 (1.1\%) \\ 
  mjson & 191 & - & 191 (71.5\%) & 73 (27.3\%) & 3 (1.1\%) \\ 
  progbase & - & 253 & 253 (94.8\%) & 11 (4.1\%) & 3 (1.1\%) \\ 
  sojo & 192 & - & 192 (71.9\%) & 72 (27\%) & 3 (1.1\%) \\ 
   \midrule
  Population & 265 & 253 & 265 (99.3\%) & 220 (82.4\%) & 58 (21.7\%)\\ 
   \bottomrule
\end{tabular}
    \vspace{0.5em}
    \caption{Observed behavior when running the \json libraries under study with  \nberroredjson \ifcorpus  files.}
    \label{tab:bench-ill-formed}
    \vspace{-1em}
\end{table}

In \autoref{tab:bench-ill-formed}, each row  summarizes the behavior of a library that processes the corpus of \ifcorpus files.
For these input files, the behavior of a library is \conform when it explicitly notifies the incorrect syntax of the input, by returning a \texttt{PARSE\_EXCEPTION} or a \texttt{NULL\_OBJECT}. A \texttt{Silent} behavior is when the library behaves as if the input was syntactically correct.
For example, the first row shows that \texttt{cookjson} is \conform for $232$ out of \nberroredjson \ifcorpus files. This library is \texttt{Silent} for $35$ files.
The last line of \autoref{tab:bench-ill-formed}, Population, indicates, in each column, the number of files in the corpus for which at least one library produces a given outcome. 

The behavior of \json libraries that process the \ifcorpus corpus is less clear-cut than the \wfcorpus. 
The share of \conform behaviors ranges from $7.9\%$ ($21$ files out of \nberroredjson) for \texttt{jjson} to $94.8\%$ ($253$ files) for \texttt{progbase}.
Meanwhile, the share of \texttt{Silent} outcome ranges from $4.1\%$ to $61.4\%$, and the share of \texttt{Error} ranges from $0\%$ for $7$ libraries up to $28.8\%$ for \texttt{jjson}.

Some libraries attempt  to build a data structure  in a ``best effort mode''. This yields a \silent behavior, which does not obviously convey the sense that the input data is ill-formed. RFC 8259 mentions  that a \textit{``\json parser MAY accept non-JSON forms or extensions''} \cite{rfc8259}. \autoref{tab:bench-ill-formed} shows that all libraries are \texttt{Silent} for some files, and $5$ libraries exhibit this behavior for more than  50\% of the ill-form files. The last line of the table also shows that the $82.4\%$ files trigger a \texttt{Silent} behavior for at least one library. This is evidence that for most files of the \ifcorpus corpus, there is at least one library that behaves as if the file was syntactically correct. On the other hand, for $99.3\%$ \ifcorpus files, at least one library correctly detects it as such.

\begin{lstlisting}[language=json, caption={Ill-formed \json Strings. Each line is extracted from a file of the Ill-formed corpus.}, label={lst:json-malformed}]
[1,]
{"Numbers cannot be hex": 0x14}
["Illegal backslash escape: \x15"]
{a:"b"}
\end{lstlisting}

\autoref{lst:json-malformed} provides examples from the \ifcorpus corpus. 
Line 1 is an array with a trailing comma. This input is interpreted as $[1]$, i.e., an array with a single value, ignoring the extra comma, by  $11$ libraries. This is a \texttt{Silent} behavior, since the behavior does not explicitly acknowledge the error in the input. The behavior of the other  $9$ libraries is \conform since $8$ of them throw an exception and \texttt{progbase} returns null. The developers of \jsonsimple implemented a test specifying how to handle this case~\cite{jsonsimpleperm}, showing that the behavior is intentional.

Lines 2, 3, and 4 of  \autoref{lst:json-malformed} illustrate other ill-formed examples that are alternatively handled in an \conform or \texttt{Silent} way. 
Line 2 is parsed silently by
$7$ libraries, which interpret the value $0x14$ as $20$. The other libraries throw an exception.
$11$ libraries  throw an exception when processing Line 3. The other implementations escape a character that should not be, and return an array containing a String.
The example at line 4 shows an object containing a non-ambiguous key without quotes. $8$ libraries still accept this input. The \orgjson test suite indicates that such a case should be handled~\cite{orgjsonperm}.
More examples in the \ifcorpus corpus trigger different behaviors, such as the acceptance of comments or the flexibility with respect to the representation of numbers.

\vspace{1em}
\begin{mdframed}[style=mpdframe]\textbf{Answer to RQ3.} 
The behavior of 13 libraries is \conform for less than $80\%$ of the \ifcorpus files. The libraries implement a \silent behavior for a large portion of the files, i.e., they decide to tolerate \ifcorpus inputs, without any explicit notification. Yet, for $99.3\%$ of these inputs, there is at least one library \conform, indicating that a multi-version \json system can increase the likelihood of a \conform behavior. 
\end{mdframed}

\subsection{\RQfour}

In this research question,  we study the behavioral diversity across all \json libraries.
First, we look at behavioral diversity  between pairs of our \nbtargetjson \json libraries. \autoref{metric:distance} (\autoref{sec:prot3}) adapts Jaccard's distance to capture the probability that a pair of libraries behaves the same for a corpus of input files. We investigate whether there are significant differences in the average pairwise distance between libraries on the \wfcorpus and the \ifcorpus corpora.

\begin{figure}[ht]
    \centering
    \includegraphics[width=0.95\columnwidth]{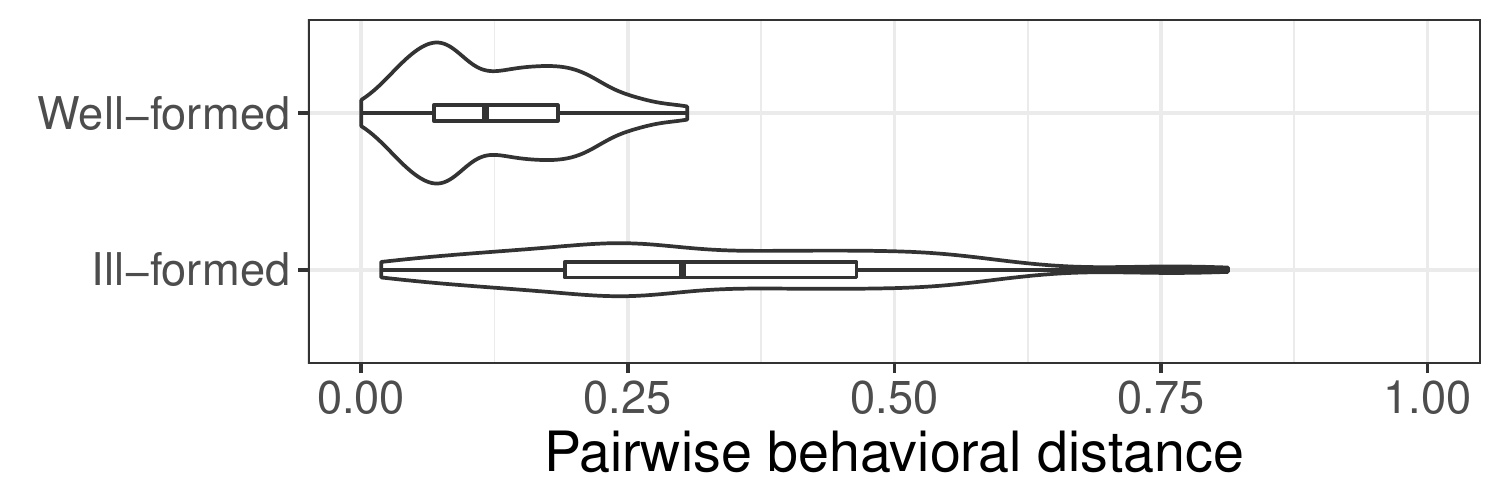}
    \caption{Distribution of pairwise behavioral distances (cf. \autoref{metric:distance}) among libraries per corpus.}
    \label{fig:pairwise-dist}
\end{figure}

\autoref{fig:pairwise-dist} shows the distributions of the pairwise behavioral distances between libraries, for the corpus of \wfcorpus and \ifcorpus files.
On the \wfcorpus corpus, the pairwise distances range from 0 (\texttt{cookjson} and \texttt{johnzon} behave exactly the same) to 0.31 (\texttt{flex-json} and \texttt{json-util} behave the same for 63 out of 206 files).
The median of the distribution is $0.12$.
For \ifcorpus files, the pairwise distances range from $0.02$ (\texttt{jackson} and  \texttt{johnzon}  behave differently on $5$/\nberroredjson files) to $0.81$ (\texttt{jjson} and \texttt{progbase} behave differently on $217$/\nberroredjson files). 
The median of the distribution is $0.30$.
Distances and variations are significantly larger on the \ifcorpus corpus. The average distance between two libraries on the \wfcorpus corpus is $0.13$, meaning that on average, two libraries yield a different outcome on $13\%$ of files. The average  distance between two libraries on the \ifcorpus corpus is $0.33$. A Welch Two Sample t-test indicates that this difference is significant with a p-value $<0.001$. 
The key observation on \autoref{fig:pairwise-dist} is that there is a larger behavioral diversity among \json libraries when they process the \ifcorpus corpus.


To consolidate our observations about behavioral diversity, we broaden the analysis to the whole set of \nbtargetjson libraries, instead of comparing pairs. 
\autoref{fig:consensus-wf} and \autoref{fig:consensus-if} show how many libraries exhibit the same behavior when processing an input \json file.

Each bar in \autoref{fig:consensus-wf} corresponds to the number of libraries that behave the same for a given share of the \wfcorpus files. A bar can have up to 3 subparts depending on the behavior that the libraries share (\conform, \silent or \error).
The x-axis ranges from $1$ to $20$: values from $1$ to $19$ correspond to the size of the subsets of libraries that have the same behavior for a file, and $20$ shows the share of files that are handled in the same way by  all libraries.
The y-axis gives the share of files that are handled in the same way by a set of libraries of a given size. The rightmost bar indicates that all \nbtargetjson libraries have a \conform behavior for $49\%$ of the \wfcorpus files ($101$ out of \nbcorrectjson). This means two things: the libraries behave the same for almost half of the inputs; the consensus is all on \conform behavior. 
The rest of the files are handled differently by subsets of the libraries. For example \textit{negative-zero.json} (line 1, \autoref{lst:json-correct}) triggers a \conform behavior for $19$ libraries and an \error for $1$, so it contributes to the bars $19$ and $1$. The bar on the left indicates that for $15.5\%$ of the files, there is a singular library with a  \silent behavior, and for $13.1\%$ there is one library with an \error behavior.
Overall, we observe that $84.9\%$ of the \wfcorpus files trigger the same  \conform behavior for at least $17$ libraries. The non-\conform behaviors are distributed among small subsets of libraries.

\begin{figure}[t]
    \centering
    \includegraphics[width=0.95\columnwidth]{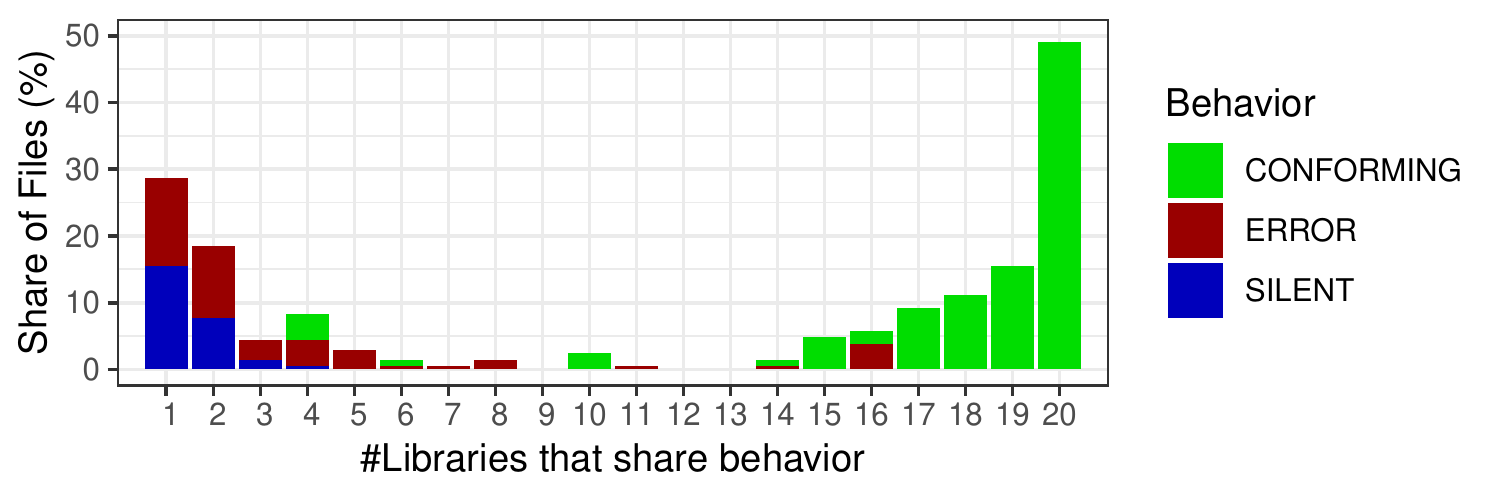}
    \caption{Distribution of the number of libraries that behave the same with the \wfcorpus corpus}
    \label{fig:consensus-wf}
    \vspace{-1em}
\end{figure}

\begin{figure}[t]
    \centering
    \includegraphics[width=0.95\columnwidth]{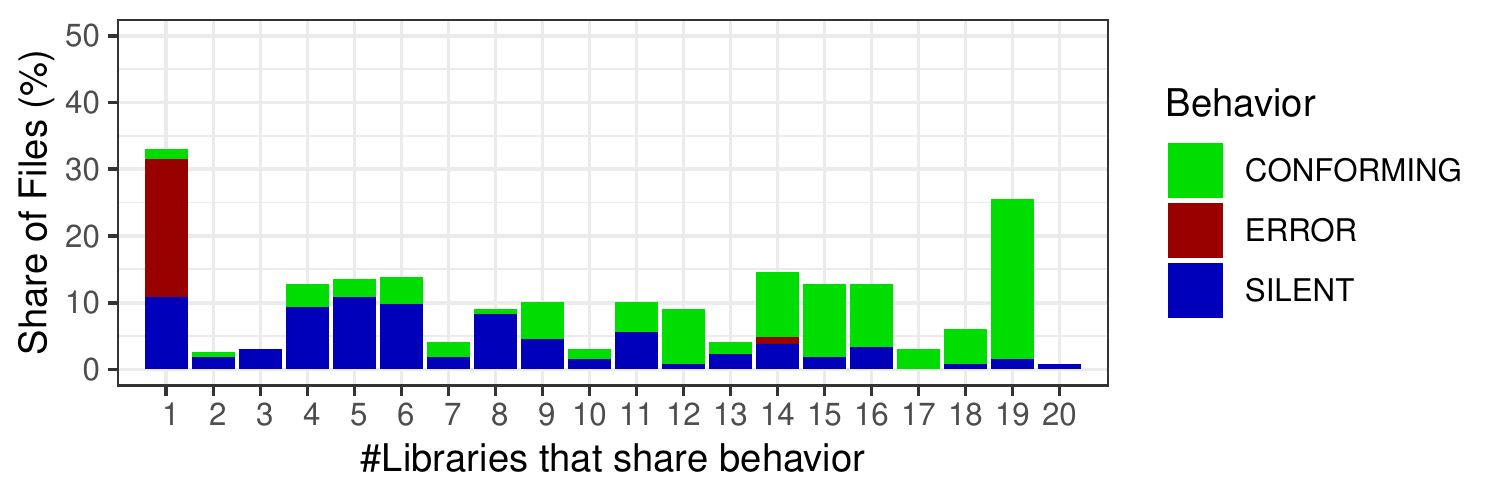}

    \caption{Distribution of the number of libraries that behave the same with the \ifcorpus corpus}
    \label{fig:consensus-if}
\end{figure}

\autoref{fig:consensus-if} shows the distribution of the number of libraries that behave the same  for  the \ifcorpus corpus. The rightmost bar indicates that only $0.75\%$ of the files ($2$ out of \nberroredjson) are processed the same way by all $20$ libraries and trigger a \silent behavior. $19$ libraries have the same \conform behavior with  $24\%$ of the files ($64$ out of \nberroredjson).
A key difference with \autoref{fig:consensus-wf} is the spread of the bars: here they are not concentrated on the extreme values but distributed in many sets of all sizes. This indicates a wider diversity of behavior.
Looking at \conform behavior, we can compare the $84.9\%$ of files that yield a \conform behavior for at least $17$ libraries in the \wfcorpus, while only $35.3\%$ of the files yield such a consensus with the  \ifcorpus corpus. Another interesting difference is the presence of \conform behavior on the left of the plot, indicating that some files are processed correctly only by one or two libraries. 


There is more behavioral diversity among libraries that process ill-formed \json files than when they process \wfcorpus files. This is consistent with the guideline from RFC 8259~\cite{rfc8259} which states explicitly that a \textit{"JSON parser MAY accept non-JSON forms or extensions."}. We observe indeed a greater probability of differing outcomes between two libraries on the \ifcorpus corpus. 

\vspace{0.5em}
\begin{mdframed}[style=mpdframe]
\textbf{Answer to RQ4.}
All libraries behave exactly the same for $49\%$ of the \wfcorpus files and 0.75\% of the \ifcorpus files. We observe a much wider diversity of behavior among \json libraries when they process \ifcorpus files.
These results show that a reliable \json processing can only be achieved through the combination of multiple libraries.
\end{mdframed}

\section{Threats to validity}
\label{sec:threats}

\textbf{Internal Validity.} 
The main internal threat to validity lies in the construction of the \json corpora. These corpora need to  cover a wide range of diverse \json inputs, and the classification between well-formed and ill-formed needs to be as accurate as possible. To limit the risks regarding the diversity of inputs, we  gather \json files from $4$ different sources, including the official \url{json.org} test suite and a test suite specifically designed to explore  the corner cases of the format. To mitigate the ambiguities between ill-formed and well-formed, we manually investigated the \json files for which a majority of libraries did not yield \conform results.  

\textbf{External Validity. } Our study is limited to \nbtargetjson Java libraries. Hence, our results might not generalize to other libraries or languages. It is important to note that the root of the observed diversity in the behavior of libraries partly comes from the \json specification itself, as well as from the difference between \json types and Java types. We believe that as long as both of these properties hold, it is likely to observe a similar diversity in other ecosystems.

\textbf{Construct validity. } In order to limit threats to construct validity, we use diverse perspectives and we do not rely on one single metric to draw the conclusion of a greater diversity of behavior among libraries on \ifcorpus inputs. We rely on the range of \conform behavior presented in RQ2 and in RQ3, while for R4 we use a notion of distance and the distribution of the number of libraries that behave the same.\looseness=-1

\section{Related Work}
\label{sec:rw}

\textbf{Analyzing behavior diversity}. Several  works study software  diversity among multiple software projects providing similar or the same functionalities. Koopman and colleagues~\cite{koopman1999comparing} propose a comparison of 13 POSIX implementations' behavior. They feed these implementations with a corpus of abnormal inputs parameters and observe the outcome. They observe that when not considering the implementation of the C library, only $3.8\%$ of failures are common to all 13 implementations.
Gashi and colleagues~\cite{gashi2004fault} examine 4 SQL server implementations and the bugs that affect them. They find that no bug affects all four implementations, and emphasize  the opportunity of building a fault tolerant system based on this diversity.
Harrand and colleagues~\cite{harrand20} study the diversity of 8 Java bytecode decompilers. They observe all decompilers do not fail with the same input files. They propose a meta decompiler that combines the results of different decompilers to build a more reliable one.
Carzaniga \cite{carzaniga2015measuring} and Gabel \cite{gabel2010study} check the input/output behavior redundancy of  code snippets with random testing. Our work contributes to this body of knowledge about natural software diversity with novel observations about \json libraries.

\textbf{Exploiting software diversity}.
Software diversity \cite{forrest1997building} has been exploited in various works for dependability, reliability, testing and  security \cite{baudry15}.
Muralidharan et al~\cite{muralidharan2016architecture} leverage code variants to adapt performance in the context of GPU code. 
Basios~\cite{basios2018darwinian} and Shacham and colleagues~\cite{shacham2009chameleon} exploit the diversity of data structure implementations to tailor the selection according to the application that uses a data structure.
Sondhi and colleagues~\cite{sondhi2019similarities} leverage similarities between library implementations to reuse test cases from one to test another. Boussaa and colleagues~\cite{boussaa2020leveraging} study family of code generators that target different languages from the same sources. They rely on metamorphic testing to automatically detect inconsistencies in these code generators.
Srivastava and colleagues~\cite{srivastava2011security} compare the multiple implementations of Java libraries to find bugs in the enforcement of security policies. 
Xu leverages the diversity of computing platforms, focusing on eight factors in OS design and implementation, to build an efficient system to detect malicious documents ~\cite{xu17}.
Our analysis of \json libraries behavior sets the foundations for future work that uses multiple implementations for a more resilient management of \json data.

\looseness=-1


    


\textbf{Analyzing \json}. The small number of existing studies that compare serialization libraries, including \json, focus on performance. To our knowledge, there is no previous work that compares functional behavior.
Maeda~\cite{maeda12} compares the performance of twelve Java serialization libraries (XML, \json and binary). Those libraries exhibit significant performance differences, while all staying in a reasonable range.
Similarly, Vanura et al~\cite{Vanura18} evaluate $49$ serialization libraries in diverse languages and propose a benchmark aiming at measuring their performance. 
In his blog post, \textit{Parsing \json is a minefield}, Seriot~\cite{parsingJSON} proposes a collection of \json files to test how \json parsers handle corner cases of the \json format. He strongly emphasizes that the specification leaves ambiguity. We integrate Seriot's collection in our corpora. 
Our work differs from these previous works as we focus on assessing and comparing the  input/output behavior of Java \json libraries.


\section{Conclusion}


The \json format has become increasingly popular in the past 20 years. The popularity of \json has fueled the development and maintenance of multiple libraries that all provide services to process \json files. While the format is thoroughly specified in RFC 8259~\cite{rfc8259}, the  specification leaves significant room for choice when implementing a specific library to process \json.
We propose the first systematic analysis of \nbtargetjson Java \json libraries. We observe that libraries make significantly different  choices of data structures to represent \json types. 
Executing the libraries on 473 \json files, we observe that the diversity of design choices is 
reflected in the input/output behavior of these libraries. Most of the libraries have a behavior \conform to the standard for \wfcorpus files, including \gson that processes $100\%$ of the files without errors.
Meanwhile, the processing of ill-formed files exhibits a significant diversity of behavior. Only $0.75\%$ of the ill-formed files are recognized as such by all libraries and all libraries exhibit non \conform behavior on some ill-formed files. Yet, when considering the collective behavior of the \json libraries, up to $99.3\%$ of the files are recognized as ill-formed by one library at least.

The essential role of \json in distributed systems calls for reliable and secure processing solutions.
Our findings open exciting possibilities in terms of software resilience, as developers who build variants of their applications with diverse \json library implementations could benefit from this natural diversity to mitigate the risks of bugs due to the mishandling of ill-formed \json data.

\section{Acknowledgement}

This work is partially supported by the Wallenberg AI, Autonomous Systems, and Software Program (WASP) funded by Knut and Alice Wallenberg Foundation and by the TrustFull project funded by the Swedish Foundation for Strategic Research.


\bibliography{main}{}
\bibliographystyle{ieeetr}

\end{document}